\documentclass[%
 reprint,
 amsmath,amssymb,
 aps,
]{revtex4-2}

\usepackage{graphicx}
\usepackage{dcolumn}
\usepackage{bm}

\begin{document}

\preprint{APS/123-QED}

\title{Third-order nonlinear femtosecond optical gating through highly scattering media}

\author{Maïmouna Bocoum}
\email{maimouna.bocoum@espci.fr}
\affiliation{%
Institut Langevin, ESPCI Paris, Université PSL, CNRS, 75005 Paris, France
}%
\author{Zhao Cheng}%
\email{zhao.cheng@ensta-paris.fr}
\author{Jaismeen Kaur}%
\email{jaismeen.kaur@ensta-paris.fr}
\author{Rodrigo Lopez-Martens}
\email{rodrigo.lopez-martens@ensta-paris.fr}
\affiliation{Laboratoire d’Optique Appliqu\'ee, CNRS, Ecole Polytechnique, ENSTA Paris, Institut Polytechnique de Paris, 181 chemin de la Huni\`ere et des Joncherettes, 91120, Palaiseau, France}%

\date{\today}

\begin{abstract}
Discriminating between ballistic and diffuse components of light propagating through highly scattering media is not only important for imaging purposes but also for investigating the fundamental diffusion properties of the medium itself. Massively developed to this end over the past 20 years, nonlinear temporal gating remains limited to $\sim 10^{-10}$ transmission factors. Here, we report nonlinear time gated measurements of highly scattered femtosecond pulses with transmission factors as low as $\approx 10^{-12}$. Our approach is based on third-order nonlinear cross-correlation of femtosecond pulses, a standard diagnostic used in high-power laser science, applied for the first time to the study of fundamental light scattering properties.
\end{abstract}

\maketitle


When an ultrashort light pulse propagates through a scattering medium, its intensity undergoes an exponential decrease with ballistic propagation quantified by the scattering coefficient $\mu_s$. Simultaneously, a slower diffused component of light rises, withholding additional information about the medium. In a transmission configuration, temporal gating of the ballistic component may be exploited for shadow imaging~\cite{hee1993femtosecond} or to simply extract of $\mu_s$ from the attenuation $e^{-\mu_s L}$ factor~\cite{tong2011measurements,wang1995true}, where $L$ is the length of the medium. In the highly scattering regime, where propagation is described by a diffusion equation~\cite{carminati2021principles}, fitting the temporal shape of either the transmitted of reflected light at longer times ($\gg$ps) provides a measure of the diffusion coefficient $D = v_e/(3\mu_s')$, where $v_e$ is the energy velocity~\cite{lagendijk1996resonant} and $\mu_s'$ the inverse of the transport mean free path~\cite{carminati2021principles}. Measuring both $\mu_s$ and $\mu_s'$ is crutial to fully characterize a scattering media as they are related by the relation $\mu_s' = \mu_s(1-g)$, where $g$ is the anisotropy which quantifies the directionality of the scattering process. Although direct measurements of $\mu_s'$ often rely on the use of coherent back-scattering (CBS) techniques~\cite{wolf1988optical,tricoli2019modeling} or photonic Ohm-law static transmission~\cite{sapienza2007observation}, time gating methods may also be applied to the characterization of  biological samples or scattering phantoms whenever the value of $v_e$ is known ~\cite{madsen1992experimental,bargigia2012time}. The main advantage of temporal gating over CBS is its sensitivity to $D$ over time as opposed to $\mu_s'$ only, hence the additional information it provides about the spatial or spectral behavior of the scatterers inside the medium. In the early 2000s, temporal gating was for instance used to demonstrate the transition from diffuse to localized propagation states when $\mu_s' \sim \lambda^{-1}$~\cite{storzer2006observation}, where $\lambda$ designates the wavelength of the scattered light. Although this interpretation has since been subject to debate and most likely attributable to fluorescence~\cite{sperling2016can}, temporal gating remains a powerful experimental tool for exploring deviations from classical diffusion behavior and their link to the mesoscopic topology of the scattering medium~\cite{van1991speed,busch1995transport,johnson2003time,wiersma2000time}. 

We often undermine how crucial the choice of temporal detection method is relative to the application or sample properties. Coherent gating either in the temporal~\cite{cobus2021transient,badon2015retrieving} or spectral domains~\cite{kop1995phase,johnson2003time} has been extensively used to probe the temporal dynamics of multiply scattered light. The measured quantity however is not the averaged diffused intensity by the medium but rather its temporal (Green function)~\cite{badon2015retrieving} or spectral response (transfer function)~\cite{kop1995phase,johnson2003time} for one realization of disorder. Probing the diffusion properties of a given scattering medium therefore requires averaging over multiple realizations of disorder~\cite{vellekoop2005determination,hee1993femtosecond,johnson2003time}. In addition, the bandwidth of the measured transfer function is limited by that of the illumination source~\cite{vellekoop2005determination}, and the maximum measurable time window either by the excursion of the delay stage used for temporal measurements or by the resolution of the spectrometer in the case of spectral measurements. This limitation is extremely problematic because non-classical propagation behavior such as localisation effects~\cite{john1987strong,segev2013anderson,schwartz2007transport} are expected for very long delays and low transmission factors. To circumvent this experimental difficulty, one must rely on incoherent temporal detection that is sensitive to the intensity of the scattered light. 

Historically, such time-resolved experiments were based on streak camera gating~\cite{yoo1990time,ho1989time,hebden1991time,bruce1995investigation}, but the linear dynamic range of digital sensors makes it inadequate for temporal acquisitions with log-variation in time. Single-photon counting detectors offer excellent sensitivity but feature limited temporal resolution, such that temporal traces have only been reported in the nanosecond range~\cite{madsen1992experimental,storzer2006observation}. In a transmission measurement where the spreading of the incident pulse scales with the Thouless time $\tau_l = L^2/D$~\cite{thouless1974electrons}, the scope of investigation is therefore restricted to highly scattering phantoms such as solid powders~\cite{watson1987searching,storzer2006observation}, unresembling biological tissues or diluted phantoms, unless the measurement is performed in (less precise) semi-infinite reflection configuration~\cite{madsen1992experimental,patterson1989time}.\\

The 1990s witnessed the emergence of nonlinear temporal gating techniques with femtosecond laser pulses~\cite{wang1991ballistic,wang1995true,tong2011measurements,mujumdar2004imaging,hauger1996time}, combining both high temporal resolution and high dynamic detection range. Although techniques such as second harmonic generation (SHG) gating~\cite{hauger1996time} or optical Kerr gating (OKG)~\cite{tong2011measurements,wang1995true,wang1991ballistic} are very efficient to probe complex media, the lowest transmission factor reported is $\sim10^{-10}$~\cite{calba2008ultrashort,tong2011measurements}, which is still too high for characterizing fat emulsions in transmission. In this work, we show how third-harmonic generation (THG), a standard technique in high-power laser science with sensitivities of $\sim 10^{-12}$ or higher, can be used to characterize a highly scattering slab in transmission. Although comparable sensitivities have been reported  using state-or-the-art setups based on optical parametric amplification or even SHG in one case~\cite{divall2004high,llereport} could in principle reach similar sensitivities, our approach offers immediate access to this record level of sensitivity using almost the highest dynamic range accessible today, all this using a commercially available device that anybody can buy and operate.  We illustrate this by performing the first simultaneous measurement of both the scattering coefficient, $\mu_s$, and reduced scattering coefficient, $\mu_s'$, of a fat emulsion in a transmission geometry.

\begin{figure}[h]
\centering\includegraphics[width=0.46\textwidth]{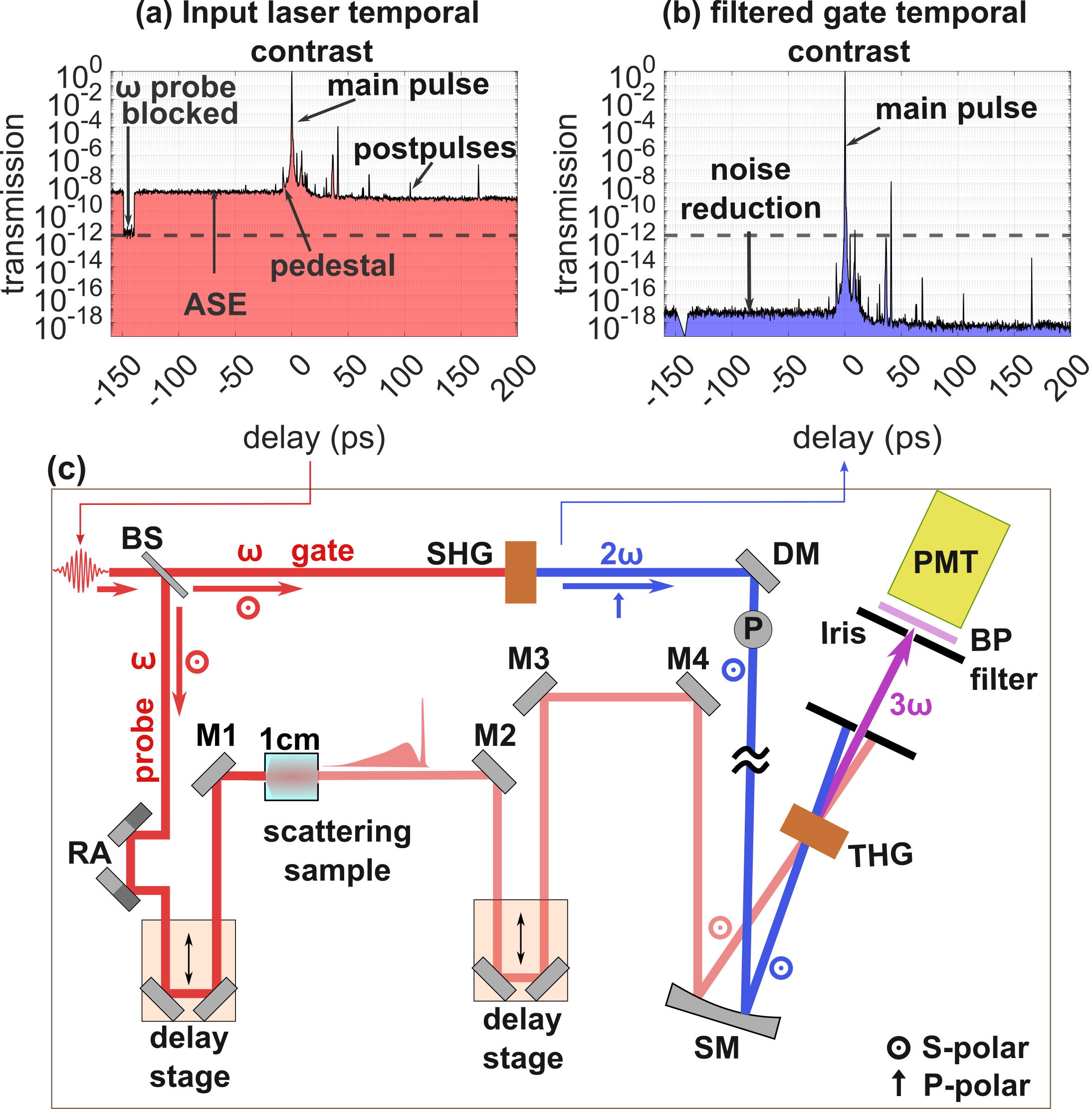}
\caption{(a) THG cross-correlation measurement of the input NIR laser pulse profile, normalized to its peak intensity. (ASE: Amplified Spontaneous Emission);(b) calculated square-root of the measured temporal NIR pulse profile (c) THG cross-correlator setup. M1,M2,M3,M4: Silver mirrors. BS: beamsplitter. RA: reflective attenuator set. DM: dichroïc mirror. P: Periscope SHG: Second Harmonic generation crystal. THG: Third harmonic generation crystal. SM: Spherical focusing mirror. Retro-reflectors are mounted on two high-stability delay stages. PMT: Photo-multiplier tube}
\label{fig:setup}
\end{figure}

THG cross-correlators were originally developed to diagnose unwanted pedestals, pre-pulses or amplified spontaneous emission (ASE) on the picosecond-to-nanosecond timescale surrounding the peak of ultra-intense femtosecond laser pulses~\cite{hong2005generation,luan1993high,itatani1998suppression,osvay2005temporal}, and can reach up to $10^{13}$ dynamic range in the near-IR (NIR) spectral region~\cite{schanz2019high,itatani1998suppression,ma2021resolving}. In our experiment, we use a commercial all-reflective third-order cross-correlator (Tundra, Ultrafast Innovations GmbH) featuring $10^{12}$ dynamic detection range. A schematic representation of the cross-correlator setup is shown in Fig~\ref{fig:setup}(c). S-polarized 30fs input pulses, centered at 790\,nm, with $400~\mathrm{\mu J}$ energy, are sent into the device at 1\,kHz repetition rate. Each pulse is separated into a probe and a gate pulse with a 5-95\%-beamsplitter. The $20~\mathrm{\mu J}$ probe pulse of $\approx 5~\mathrm{mm}$ in diameter is attenuated using variable calibrated reflectivity mirrors (RA in Figure~\ref{fig:setup}) so as to keep the PMT response linear. The pulse is then sent through the scattering medium, located in between two long-range delay lines, while the gate pulse undergoes type I SHG in a BBO crystal to generate a P-polarized SHG gate pulse centered at $2\omega$, where $\omega$ is the central laser frequency. The SHG gate pulse is filtered out from the residual NIR pulse using dichroic mirrors, converted back to S-polarization with a periscope and mixed with the time-delayed S-component of the scattered pulse in a type I THG crystal to generate the cross-correlation signal at $3\omega$. The cross-correlation trace is obtained by recording the spatio-spectrally filtered $3\omega$ signal with a solar blind photo-multiplier tube (PMT) as a function of delay between scattered and gate pulses, with a maximum delay of up to $\pm 2$\,ns and $\sim100$\,fs temporal resolution at best. Each data point is an average of 100 consecutive shots recorded in 100\,fs time steps, except for the range spanning from -10\,ps to +10\,ps, where data are recorded in 10\,fs time steps. A typical cross-correlation trace of the (unscattered) input laser pulse is shown in Figure~\ref{fig:setup}(a) over a 350\,ps time window around the pulse peak. A detection noise floor of $\sim10^{-12}$ can indeed be measured by blocking the $\omega$ probe arm at early times in the trace. The key to noise reduction in THG cross-correlators is the very low level of self-generated signal leaking from either arms of the optical setup into the $3\omega$ detector~\cite{luan1993high,schanz2019high}.

\begin{figure}[h]
\centering\includegraphics[width=0.5\textwidth]{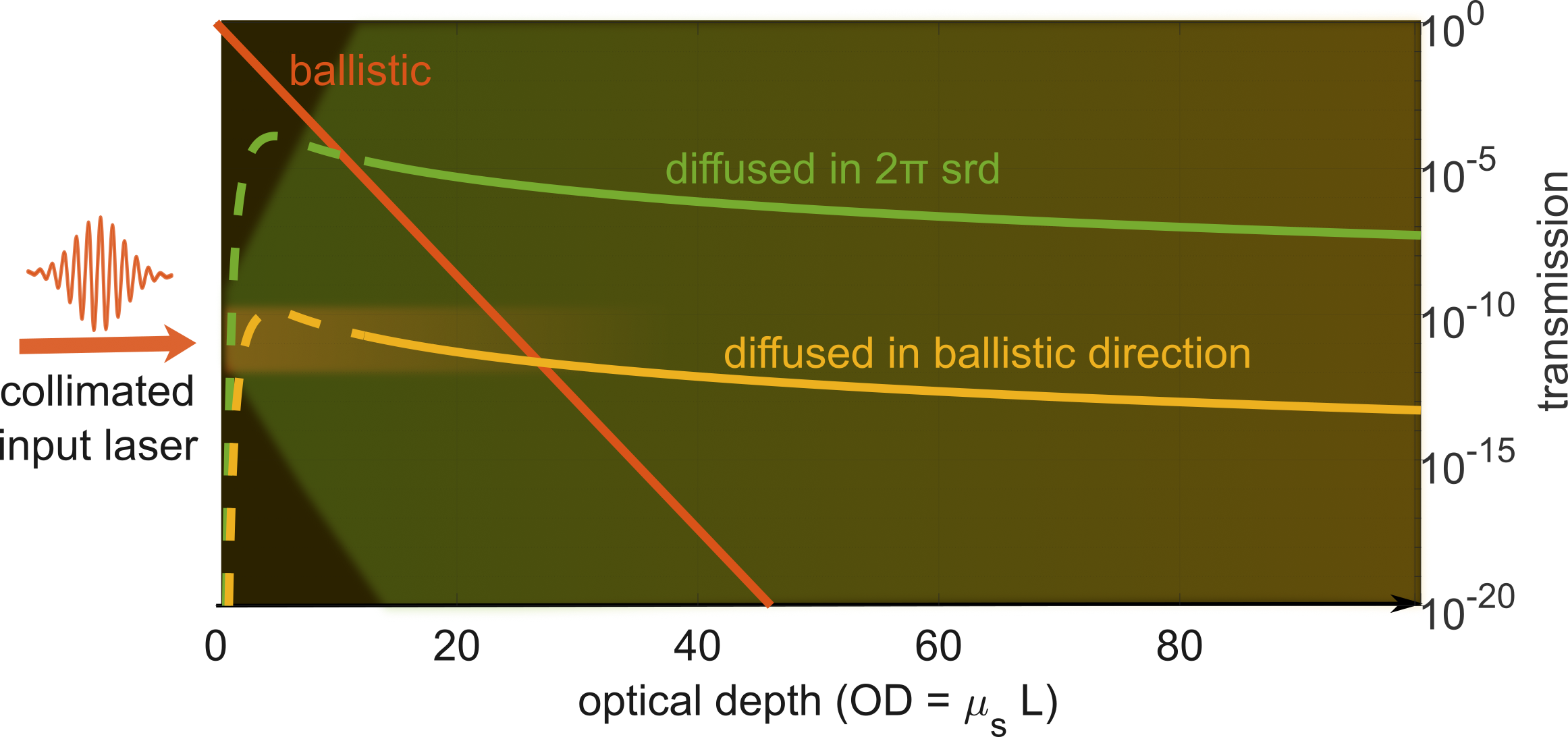}
\caption{Maximum transmission plotted as a function of $L$ for the ballistic component (red curve) and the diffused light collected over $2\pi~\mathrm{srd}$ (green upper curve), or limited to $2\pi 10^{-6}~\mathrm{srd}$ in the ballistic direction (yellow lower curve) using a $\tau = 25~\mathrm{fs}$ temporal gate. Dash lines indicates maximum transmission occurs at a time $t < 10 (v_e\mu_s')^{-1}$, so as to outline the limited accuracy of equation~\ref{eq:scattering}. the scattering properties of the medium are set to $\mu_s=20~\mathrm{cm^{-1}}$ , $\mu_a =0~\mathrm{cm^{-1}}$ and $g=0.5$}
\label{fig:transmission}
\end{figure}

To demonstrate the potential of a $\sim 10^{-12}$ sensitivity for the detection of diffused light, we characterized the scattering properties of a commercial intralipid-10\% emulsion used in previous scattering experiments~\cite{bocoum2018two}. Fat emulsions are extensively used as light scattering models or as phantoms for biological applications~\cite{michels2008optical,bocoum2018two}. The reason is an accessible price, low absorption ($\mu_a \ll \mu_s$), scalability of scattering properties with dilution, and the spherical shape of the fat droplets, which makes them easy to model using Mie theory. Despite this, measuring their scattering properties remains a challenge and different methods can yield significantly different results~\cite{pifferi2005performance}. In particular, temporal gating characterization requires fulfilling the diffusion approximation, ie (i) $ L\mu_s'\gg 1$, (ii) $\mu_a \ll \mu_s$ and (iii) $t\gg (v_e\mu_s')^{-1}$. Considering a short pulse incident on a fat emulsion slab modeled by a Dirac function in time, the average intensity $I_d[\mathrm{W/cm^2}]$ of the diffused component writes~\cite{carminati2021principles}:
 
\begin{equation}
\label{eq:scattering}
I_d(t)= E_{\mathrm{in}}\frac{H(t) D}{d}\sum_{m=1}^{\infty}\frac{m\pi}{d} \sin(\frac{\pi m L}{d})e^{-\frac{\pi^2m^2Dt}{L^2}}e^{-\mu_a c t},
\end{equation}

\noindent where $H(t)$ is the Heaviside function, $t$ the time following the pulse arrival time on the slab, $L$ the slab width, $d =L + z_0$, with $z_0$ as an adjustable parameter on the order of $z_0\sim (\mu_s')^{-1}$, necessary to account for boundary conditions at the slab interface~\cite{carminati2021principles}, and $E_{\mathrm{in}}[\mathrm{J/cm^2}]$ the incident pulse fluence. We now define the transmission, $T_d$, as the ratio between detected and incident pulse fluence, over the gate time $\tau$, the solid angle $\delta \Omega$ and a single polarisation state, such that $T_d =  \tau \delta \Omega I_d(t)/(4 \pi E_{\mathrm{in}})$. Recalling that for non-resonant scattering media, $v_e = c/n$, where $c$ is the velocity of light in vacuum, and $n$ the effective index of the medium~\cite{carminati2021principles}, the maximum transmission evaluated from Eq~\ref{eq:scattering} occurs at $t\sim 0.09 L^2/D$, and yields the scaling law:

\begin{equation}
\label{eq:scaling}
T_{d,\mathrm{max}} \sim  \frac{0.1 \pi c \tau \delta \Omega }{n(1-g)^2\mu_s^2 L^3} \sim 1.4 \frac{10^{-9}}{(1-g)^2 OD^2 L(cm)}, 
\end{equation}

\noindent where the numerical evaluation on the right hand side is done for $\tau = 25~\mathrm{fs}$, $n=1.33$ and $\delta \Omega = 2\pi 10^{-6}~\mathrm{srd}$, taken as a reference value from the SHG gating measurement reported in~\cite{calba2008ultrashort}. This simple scaling shows how crucial the detection sensitivity should be in order to explore large optical depths. In Fig~\ref{fig:transmission}, we plot the maximum transmission of the diffused component obtained using Eq~\ref{eq:scattering} as we increase the propagation length $L$, for a given value $\mu_s=20~\mathrm{cm^{-1}}$ and $\mu_a =0~\mathrm{cm^{-1}}$ and $g=0.5$. The ballistic transmission in the same condition is plotted in red for comparison. In particular, by imposing $L\mu_s'\ge 10$ to satisfy the diffusion approximation, we have $T_{d,\mathrm{max}}\lesssim  10^{-11}/L(cm)$, which is lower than the lowest transmission factor measured in~\cite{calba2008ultrashort,tong2011measurements}, unless phantoms thinner than $\sim 1\mathrm{mm}$ are used. In particular, the simultaneous measurement of both $\mu_s$ and $\mu_s'$ in fat emulsions has never been done using non-linear gated detection, but rather restricted to phantoms with high $\mu_s'$ and low $L$ such as TiO$_\mathrm{2}$ powders. 

\begin{figure}
\centering\includegraphics[width = 0.47\textwidth]{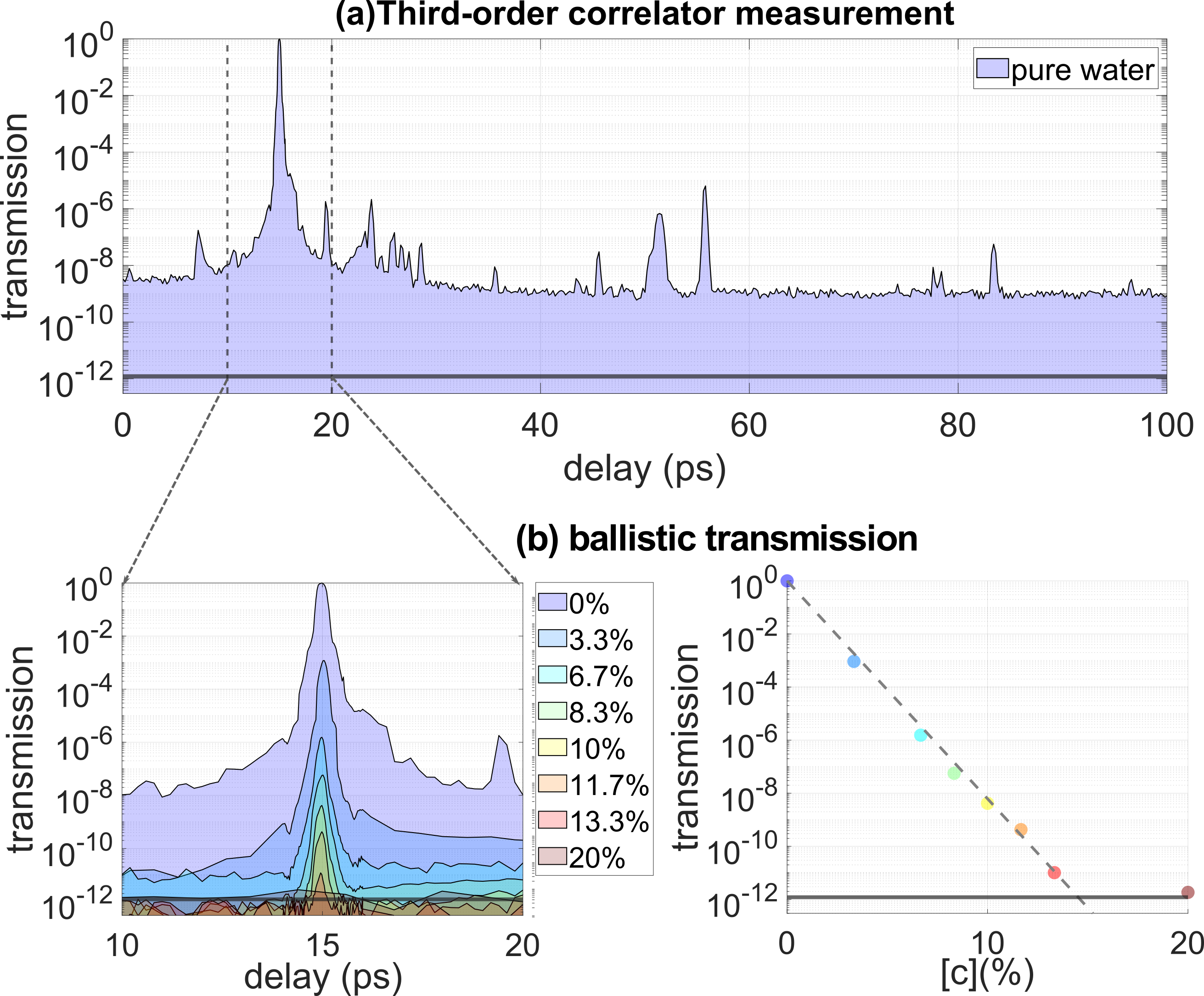}
\caption{(a) THG cross-correlation measurement through the cell filled with pure demineralized water and used for reference (b) Same measurement with increasing intralipid-10\% concentration [c] (color coded) zoomed in over a 20\,ps time window around the ballistic component centered at $\sim~ 15~\mathrm{ps}$ (left) and linear fit of the peak value using $\mu_{s} = (189\pm 10)[c]~\mathrm{cm^{-1}}$ (right).}
\label{fig:ballistic}
\end{figure}

In our experiment, we measured the temporal transmission through a $L=10~\mathrm{mm}$ long PMMA cuvette designed for absorption spectrometers (@plastibran) filled with intralipid-10\% solution diluted in water at varying concentration $[c]$. The reference trace shown in Fig~\ref{fig:transmission}(a) corresponds to a cuvette filled with only pure demineralized water and used for normalisation. The ballistic peak is measured at a delay of $+15\pm 0.2$ps, corresponding to the propagation delay induced by the water-filled cuvette relative to air (see Figure~\ref{fig:setup}(a)). Note that we expect a change in effective index limited to $\delta n\sim4.10^{-4}$ for the highest concentration of intralipid, which means a ballistic delay shift hardly resolvable with our detector. On the left panel of Fig~\ref{fig:ballistic}(b), a zoom on the ballistic component is plotted for $[c] = 3.3, 6.7, 8.3,10,11.7$ and $20\%$, and the peak value is reported on the right panel plot using the same color scale. A good fit of the ballistic attenuation is extracted from the $[c]= 13.3$\% trace and represented by the black dotted line. we retrieve $\mu_{s} = (189\pm 10)[c]~\mathrm{cm^{-1}}$, where $0\le[c]\le 1$ is the dimensionless diluted concentration of intralipid. This value is lower than $\mu_{s,th} = 281[c]~\mathrm{cm^{-1}}$ reported in~\cite{van1991light} at $790~\mathrm{nm}$, but higher than $\mu_{s,th} = 100[c]~\mathrm{cm^{-1}}$ reported in~\cite{michels2008optical}. Although the use of different brands or preparation methods could explain the discrepancy with values found in literature~\cite{michels2008optical}, we believe our method to be more precise because it is based on temporal gating. 

\begin{figure}
\centering\includegraphics[width = 0.5\textwidth]{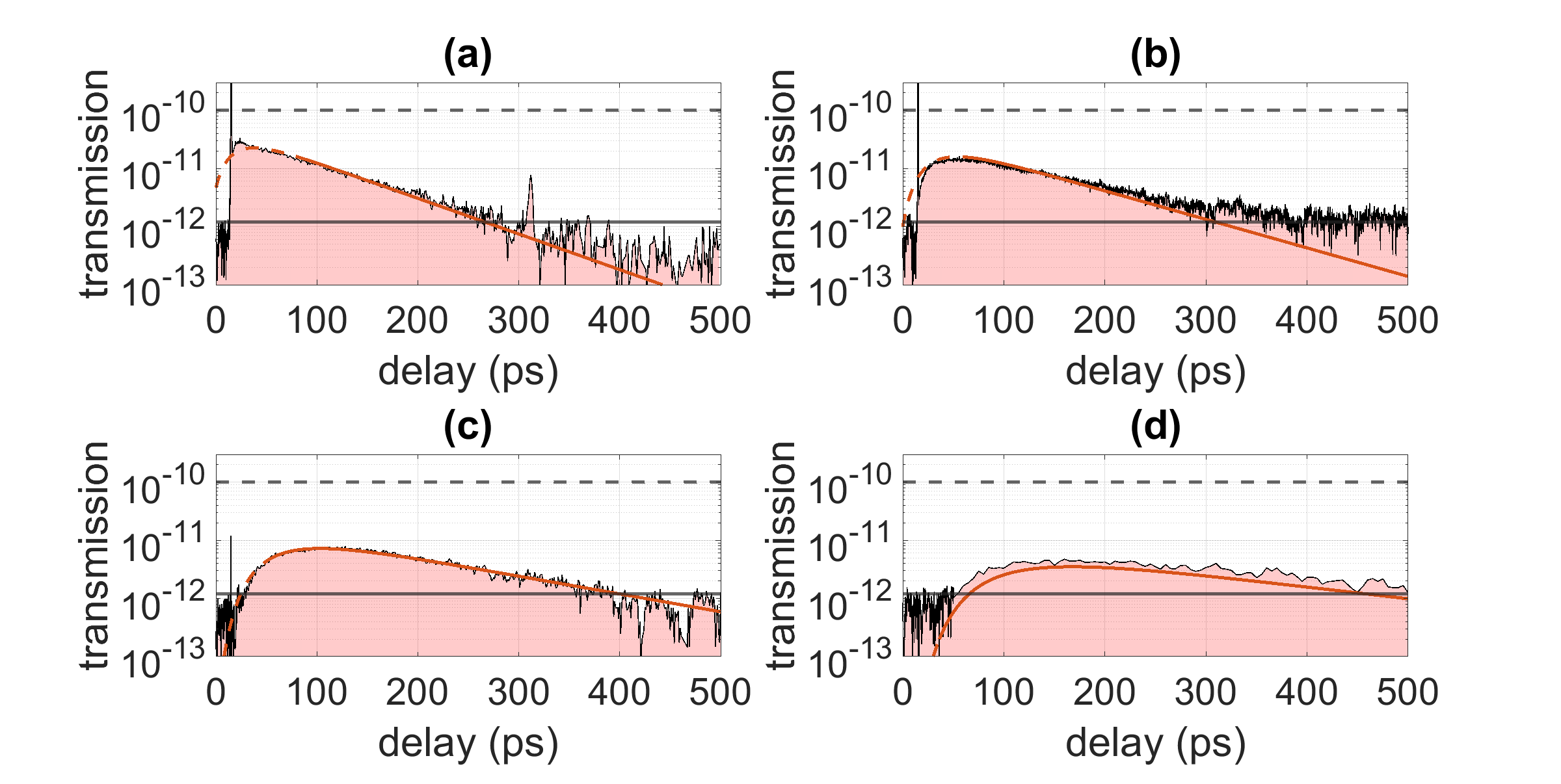}
\caption{THG cross-correlation traces obtained for [c] = 6.7\% (a), [c] = 8.3\% (b),[c] = 13.3\% (c) and [c] = 20\% (d). All red line fits are extracted from (c) providing $\mu_s' = (79\pm 7)[c]~\mathrm{cm^{-1}}$ and based on Eq~\ref{eq:scattering} with $\tau=25~\mathrm{fs}$, $\delta \Omega = 5.1\pi 10^{-6}~\mathrm{srd}$. The horizontal solid and dotted grey lines correspond respectively to our detector noise level and the lowest reported in the literature~\cite{tong2011measurements,calba2008ultrashort}}
\label{fig:diffuse}
\end{figure}

\noindent The value $\mu_s'$ is obtained by fitting the temporal profile at long times with Eq~\ref{eq:scattering}, using the same trace used to measure $\mu_s$. The effective index $n$ is extrapolated from the relative percentage of water and soybean at a given concentration $[c]$~\cite{michels2008optical}, $\tau = 25fs$, and by manually adjusting the angle of collection to $\delta \Omega = 5.1\pi 10^{-6}~\mathrm{srd}$, corresponding to an half collection angle of $\sim 0.13$°. We obtained the fit plotted in red in Fig~\ref{fig:diffuse}(c), where we retrieve $\mu_s' = (79\pm 7)[c]~\mathrm{cm^{-1}}$. Note that at this concentration, the diffusion approximation is verified since $\mu_s'L \gtrsim 10$. By fixing that value of $\mu_s'$, we superimposed theoretical prediction with experimental data obtained for $[c] = 6.7, 8.3$ and $20\%$ in Fig~\ref{fig:diffuse}(a,b and d), respectively. The slight departure from the theoretical fit observed in Fig~\ref{fig:diffuse}(a) and Fig~\ref{fig:diffuse}(b) correspond to an error of up to $13\%$ in the evaluation of $\mu_s'$. This slope deviation may be attributed to error bars in the concentration calibration of our samples, or to a slight departure from the diffusion approximation, since in both cases, $\mu_s'L < 10$. In particular, the residual presence of ballistic post-pulses at long delays (e.g. the peak observed around +310\,ps in Fig~\ref{fig:diffuse}(a)) indicate that we are not in a purely diffusive regime. The predicted temporal shape fits much better at $[c]=20\%$ shown in Fig~\ref{fig:diffuse}(d), as we would expect. However, we do observe an unexpected $\sim30$\% higher transmission that cannot be explained by considering a variation in the average index of the medium. Measuring such features was clearly out of reach for the most sensitive nonlinear gating detection methods reported thus far~\cite{tong2011measurements,calba2008ultrashort}, and indicated in back dotted line in all caption of Fig~\ref{fig:diffuse}. To ensure the reproducibility of our results, we repeated the measurement multiple times over several days and got the same result within 5\% error. Finally, from the combined measurement of both $\mu_s$ and $\mu_s'$, we estimate the anisotropic factor $g = 1 - \frac{\mu_s'}{\mu_s} = 0.58\pm 0.08$. A theoretical value of $g_{th} =0.64$ is predicted in~\cite{van1991light} and $g_{th} =0.32$ in~\cite{michels2008optical} at $790$nm wavelength.\\

In conclusion, we used THG cross-correlation to perform the first simultaneous measurement of both $\mu_s$ and $\mu_s'$ by measuring the ballistic transmission and long time semi-log variation of a short femtosecond NIR pulse transmitted through a scattering solution of [c]-diluted solution of intralipid-10\% in water. We measured $\mu_s = (189\pm 10)[c]~\mathrm{cm^{-1}}$ and $\mu_s' = (79\pm 7)[c]~\mathrm{cm^{-1}}$, from which we deduce $g = 0.58\pm 0.08$ at $790$nm wavelength. Although intralipid is extensively used in scattering experiments and in the design of biological phantoms, the simulateneous measurement of both $\mu_s$ and $\mu_s'$ had never been performed using nonlinear temporal gating. The reason for this is two fold: (i) nonlinear temporal gating is highly restricted in angular collection angle such that expected transmissions can easily be $\le 10^{-10}$ for diffused light, and (ii) all detection systems used so far were limited in dynamic range to $\sim10^{10}$, making the characterization of the diffused light almost impossible. We have demonstrated how to overcome these challenges by implementing third-order nonlinear femtosecond temporal gating with a limit of detection of the order of $\sim 10^{-12}$, which is two orders of magnitude beyond the current state-of-the-art in terms of time-gated light scattering measurements. The sensitivity of most recent commercially available THG cross-correlators can reach up to $\sim 10^{-14}$ (https://www.ultrafast-innovations.com/devices/TUNDRA.html
). This record level of sensitivity could greatly facilitate the characterization of fat emulsion phantoms and open the door to the exploration of diffusion dynamics that may deviate from classical prediction.

\begin{acknowledgments}
We thank Romain Pierrat, Arthur Goetschy and Remi Carminati for useful discussions.
\end{acknowledgments}

\appendix

\nocite{*}

\bibliography{sample.bib}

\end{document}